\documentclass[superscriptaddress,twocolumn,aps,prd,showpacs,floatfix,10pt,nofootinbib]{revtex4-1}
\usepackage{amssymb,amsmath}
\usepackage[dvips]{graphics,color}
\usepackage[colorlinks,hyperindex]{hyperref}
\usepackage{epsfig}\usepackage{float,stmaryrd,textcomp,bm,mathbbol}\usepackage{float,xcolor,upgreek,yfonts,tikz}
\usepackage{graphicx}

\newcommand{\ba}{\begin{eqnarray}}
\newcommand{\ea}{\end{eqnarray}}
\newcommand{\enge}{\end{equation}}

\newcommand{\beq}{\begin{eqnarray}}
\newcommand{\benu}{\begin{enumerate}}
\newcommand{\enu}{\end{enumerate}}
\newcommand{\eeq}{\end{eqnarray}}

\hypersetup{hidelinks,backref=true,pagebackref=true,hyperindex=true,colorlinks=true,breaklinks=true,urlcolor= blue}
\hypersetup{%
  colorlinks = true,
  linkcolor  = blue,
    citecolor = cyan,
}
\newcommand\orcidroldao{{\href{https://orcid.org/0000-0003-3978-532X}{\orcidicon}}}
\newcommand{\orcidicon}{%
	\begin{tikzpicture}
	\draw[lime, fill=lime] (0,0)
		circle [radius=0.16]
		node[white] {{\fontfamily{qag}\selectfont \tiny ID}};
	\draw[white, fill=white] (-0.0625,0.095)
		circle [radius=0.007];
	\end{tikzpicture}	\hspace{-2mm}
}

\definecolor{purple}{rgb}{1,0,1}
\definecolor{lime}{HTML}{A6CE39} 

 \newcommand{\bea}{\begin{eqnarray}}
 \newcommand{\eea}{\end{eqnarray}}

\usepackage{hyperref}
\usepackage{xcolor}

 \newcommand{\bltx}{\textcolor{black}}
 \newcommand{\pptx}{\textcolor{black}}

\begin{document}

\title{AdS$_5$--Schwarzschild deformed black branes and hydrodynamic transport coefficients}

\author{A. J. Ferreira--Martins}
\email{andre.juan@ufabc.edu.br}
\affiliation{CCNH, Universidade Federal do ABC - UFABC,  09210-580, Santo Andr\'e, Brazil.}
\author{P. Meert}
\email{pedro.meert@ufabc.edu.br} 
\affiliation{CCNH, Universidade Federal do ABC - UFABC,  09210-580, Santo Andr\'e, Brazil.}
\author{R. da Rocha\orcidroldao\!\!}
\email{roldao.rocha@ufabc.edu.br}
\affiliation{CMCC,  Federal University of ABC, 09210-580, Santo Andr\'e, Brazil.}


\begin{abstract}
A family of deformed AdS$_5$--Schwarzschild black branes is here derived, employing the membrane paradigm of AdS/CFT. \textcolor{black}{The solution of the Einstein--Hilbert action, with the Gibbons--Hawking term and a counter-term that eliminates eventual divergences, yields a partition function associated to the dual theory which allows the computation of the entropy, pressure and free energy, as state functions, in the canonical ensemble. AdS/CFT near-horizon methods are then implemented to compute the shear viscosity-to-entropy ratio, then restricting the range of the  parameter that defines a family of deformed black branes.} \end{abstract}
\maketitle

\section{Introduction}

AdS/CFT is a paradigm relating gravity in anti-de Sitter (AdS) spacetime to a large-$N$ conformal field theory (CFT), located on the AdS codimension-1 boundary. Perturbatively, considering an $1/N$ expansion, quantum fields in the bulk correspond to CFT operators \cite{rangamani, minwalla, hubeny_hard}. The dynamics of Einstein's equations, describing weakly coupled gravity in an AdS space, rules the corresponding dynamics of the energy-momentum tensor of strongly coupled QFTs on the AdS boundary. In the $N\to\infty$ t' Hooft regime, keeping a fixed coupling, the gauge theory on the boundary is an effective classical theory.

The AdS boundary is usually identified to a 4D brane. Braneworld models describe a brane that has tension, $\sigma$, constrained to both the bulk and the brane cosmological 
constants \cite{Casadio:2015gea,daRocha:2017cxu}. General relativity (GR) describes gravity in an infinitely rigid brane, with an infinite tension. However, recent works derived a strong bound for the  finite brane tension, lying in the bound $\sigma \gtrsim  2.81\times10^{-6} \;{\rm GeV^4}$ \cite{Casadio:2016aum,Fernandes-Silva:2019fez}.  This condition in fact produces a physically correct low energy limit, allowing the construction of an AdS/CFT membrane paradigm analogue of any classical GR solution \cite{ads_memb,first,mgd1,mgd2,daRocha:2017cxu,Casadio:2015gea,maartens}.  One can also describe the AdS bulk gravity by a black hole, which behaves as a fluid at its own horizon, in the membrane paradigm.
Einstein's equations near the horizon of the black hole reduce to the Navier-Stokes equations for the fluid \cite{rangamani, minwalla, hubeny_hard}. A fluid at the black hole horizon mimics a fluid at the AdS boundary \cite{Eling:2009sj,Antoniadis:1990ew,Antoniadis:1998ig,ads_memb}, introducing an useful dictionary, linking brane models and the membrane paradigm of AdS/CFT. Here we aim to derive \textcolor{black}{new deformed asymptotically AdS black branes} and use the shear viscosity-to-entropy density ratio, $\frac{\eta}{s}$, and the deformed black brane temperature, to impose viscosity bounds to the free parameter in these new solutions.
\textcolor{black}{In the context of AdS/CFT correspondence, a precise relationship between the gravitational result and the dual field theory is then established, and further discussed.}

\textcolor{black}{In AdS/CFT, the AdS$_5$-Schwarzschild black brane is dual to the gauge theory describing the strongly-coupled, large-$N_c$, $\mathcal{N}=4$, plasma. In this scheme, the famous ratio $\frac{\eta}{s} = \frac{1}{4\pi}$ (and the conjectured KSS bound) is obtained, which is indeed a quite small value, compared to ordinary materials. However, if large-$N_c$ gauge theories considered by AdS/CFT are good approximations to QCD, one could expect that this result may be applied to the quark-gluon plasma (QGP) \cite{qgp}. In fact, experiments in the Relativistic Heavy Ion Collider (RHIC) have shown that the QGP behaves like a viscous fluid with very small viscosity, which implies that the QGP is strongly-coupled, which discards the possibility of using perturbative QCD to the study of the plasma \cite{qgp_exps}. Thus, AdS/CFT may present itself as an alternative to the QGP research and generalizations thereof \cite{qgpexp1, qgpexp2}.}

Previously, we have explored the technique employed here to derive a family of solutions that consists of a deformation in the AdS$_4$--Reissner--Nordstr\"om background, and its potential applications to AdS/CMT \cite{Ferreira-Martins:2019wym}. By embedding the brane into a higher dimensional bulk, we were able to mimic the Hamiltonian and momentum constrains from the ADM formalism for static configurations of the metric field \cite{Casadio:2001jg,Abdalla:2006qj}. These equations turn out to be a weaker condition on the metric functions, allowing for a family of deformations of solutions from classical GR. In the present work we apply a similar procedure to the AdS$_5$--Schwarzschild black brane \cite{BoschiFilho:2004ci,BoschiFilho:2003zi}.

The paper is organized as follows: in Sect. \ref{sec:lrt}  the relevant results of linear response theory and fluid dynamics are briefly presented within the hydrodynamics formalism, followed by a presentation of the AdS/CFT duality. Sect.  \ref{sec:eta_s_sads_5} is then devoted to derive the AdS$_5$--Schwarzschild deformed gravitational background. \textcolor{black}{The solution of the Einstein--Hilbert action, also containing  the Gibbons--Hawking term and a counter-term that precludes divergences, yields a partition function for the dual theory. Hence, entropy, pressure and free energy, are computed as state functions, in the canonical ensemble. The   explicit computation of the $\frac{\eta}{s}$ ratio is carried out for the family AdS$_5$--Schwarzschild deformed  black branes in} Sect. \ref{sec:eta_s_ads_mgd_5}. The saturation of $\frac{\eta}{s}$ and the black brane temperature therefore is shown to constrain the free parameter AdS$_5$--Schwarzschild deformed  black brane, driving the family of deformed branes to two unique solutions: the standard AdS$_5$--Schwarzschild black brane and a new black brane solution. The concluding remarks are then presented in Sect. \ref{sec:5}.

\section{Hydrodynamics and linear response theory}
\label{sec:lrt}

\par The so called hydrodynamic limit is characterized by the long-wavelength, low-energy regime \cite{Rangamani:2009xk}, and is often applicable to describe conserved quantities. As an effective description of field theory, hydrodynamics naturally does not contain the details of a microscopic theory. These are encoded into the transport coefficients, among which the shear viscosity, $\eta$, plays a prominent role.

\par The macroscopic variables encoded in the energy-momentum stress tensor, $T^{\mu \nu}$, along with its conservation law, $\partial_\mu T^{\mu \nu} = 0$, describe a simple fluid. In general, one introduces a constitutive equation by determining the form of $T^{\mu \nu}$ in a derivative expansion, given in terms of the normalized fluid velocity field $u^\mu(x^\nu)$, its pressure field $p(x^\mu)$ and its rest-frame energy density $\rho(x^\mu)$.

To first order in the derivative expansion, the stress tensor is expressed as \cite{Rangamani:2009xk, hubeny_hard}
\begin{equation} \label{TEM:diss}
T^{\mu \nu} = p\left(\eta^{\mu \nu} + u^\mu u^\nu  \right ) + \rho u^\mu u^\nu  + \tau^{\mu \nu} \ ,
\end{equation}
where $\tau^{\mu \nu}$, the term which is first-order in derivatives, carries dissipative effects. The constitutive equation for a viscous fluid, as defined above, yields both the continuity and Navier--Stokes equations.  For a theory described by an action functional $S$, the coupling of an operator $\mathcal{O}$ to an external source $\upvarphi^{(0)}$ reads \cite{son_hydro}
\begin{equation}
	S \mapsto S + \int \mathrm{d}^4x \upvarphi^{(0)}(t, \bm{x}) \mathcal{O}(t, \bm{x})\ . 
\end{equation}
One is often interested in determining the response in $\mathcal{O}$, which, up to first order in $\upvarphi^{(0)}$, is known as linear response theory. The one-point function reads \cite{son_hydro}
\begin{equation}
\updelta \left \langle \mathcal{O} (\omega, \bm{\mathfrak{q}})\right \rangle = - G_R^{\mathcal{O}, \mathcal{O}} (\omega, \bm{\mathfrak{q}}) \upvarphi^{(0)}(\omega, \bm{\mathfrak{q}}) \ ,
\label{eq:response_def}
\end{equation}
\noindent where $G_R^{\mathcal{O}, \mathcal{O}} (\omega, \bm{\mathfrak{q}})$ is the retarded Green's function \cite{natsuume}.   The response of $\tau^{\mu \nu}$ under gravitational fluctuations is determined by an off-diagonal perturbation term, $h_{xy}^{(0)}$, leading to the perturbed metric \cite{minwalla, hubeny_hard}:
\begin{equation}
g_{\mu \nu}^{(0)}\mathrm{d}x^\mu \mathrm{d}x^\nu = \eta_{\mu \nu} \mathrm{d}x^\mu \mathrm{d}x^\nu + 2h_{xy}^{(0)}(t) \mathrm{d}x \mathrm{d}y,
\label{eq:perturbed_metric}
\end{equation} 
 yielding the response  \cite{son_hydro}
\begin{equation}
\updelta \left \langle \tau^{xy} (\omega, \bm{\mathfrak{q}} = \bm{0}) \right \rangle = i \omega \eta h_{xy}^{(0)}=- G_R^{xy, xy} h_{xy}^{(0)}\ . 
 \label{oooo}
\end{equation}
and  the Kubo formula
\begin{equation}
\eta = - \lim_{\omega \rightarrow 0} \frac{1}{\omega} \operatorname{Im}\left (G_R^{xy, xy} (\omega, \bm{0}) \right ). \label{xxx}
\end{equation}
\noindent Computation of the retarded Green's function is straightforwardly achieved, once the GKPW relation \cite{gkp1,gkp2} is regarded.
It yields the following expression for the one-point function, \cite{gkp2, Witten:1998qj}, 
\begin{equation}
\left \langle \mathcal{O} \right \rangle_S = \frac{\updelta \bar{S}[\upvarphi^{(0)}]}{\updelta \upvarphi^{(0)}} \ .
\label{eq:one_point}
\end{equation}
\noindent One considers the bulk theory to be GR, with negative cosmological constant, $\Lambda_5$. Therefore the action reads 
\begin{equation}\label{eq:fullAction}
	S=\frac{1}{16\pi}\int d^{5}x\sqrt{-g}\left(R-2\Lambda_5\right)+S_{mat} \ ,
\end{equation}
 \noindent where $S_{mat}$ is specified by the boundary theory of interest. The action for massless scalar field is just a kinetic term. 
A particular case of interest is the AdS$_5$--Schwarzschild spacetime,
\begin{equation}
ds^2 = -\frac{r_0^2}{u^2} f(u) \mathrm{d}t^2 + \frac{1}{u^2 f(u)} \mathrm{d}u^2 + \frac{r_0^2}{u^2} \delta_{ij} \mathrm{d}x^i \mathrm{d}x^j \ ,
\label{eq:sads_5_u}
\end{equation}
\noindent where $f(u) = 1-u^4$, with \bltx{$u=r_0/r$ defining the radial coordinate hereon in this paper, where $r_0$ is the horizon radius}. Hence $u=1$ locates the horizon, whereas $u=0$ is the spacetime  boundary.  For $u\rightarrow 0$, Eq. \eqref{eq:sads_5_u} reads
\begin{eqnarray}
ds^2= \frac{r_0^2}{u^2} \left (-\mathrm{d}t^2 + \frac{1}{r_0^2}\mathrm{d}u^2 + \delta_{ij} \mathrm{d}x^i \mathrm{d}x^j \right ) \ .
\label{eq:asym_ads}
\end{eqnarray}
The one-point function, Eq. \eqref{eq:one_point}, depends only on the matter contribution when computing the on-shell action. Assuming $\upvarphi = \upvarphi(u)$, \bltx{and denoting by a dot the derivative with respect to $u$}, the action for the massless scalar field at the boundary becomes
\begin{eqnarray}
\!\!\!\!\!\!\!\!\!\!\!S \!\sim\! \int\! \mathrm{d}^4x\! \left .\left ( \frac{r_0^4}{2u^3} \upvarphi \dot\upvarphi\right )\right \rvert_{u=0} \!\!\!\!\!+ \int \mathrm{d}^5x \!
\left (\frac{r_0^4}{2u^3} \ddot\upvarphi \!-\!\frac{3r_0^4}{2u^4} \dot\upvarphi\right )\! \upvarphi.
\label{eq:onshell_quase}
\end{eqnarray}
Eq. (\ref{eq:onshell_quase}) is just the EOM for the scalar field, whose asymptotic solution reads
\begin{equation}
 \upvarphi \sim \upvarphi^{(0)} \left (1 + \upvarphi^{(1)} u^4 \right ) \ .
 \label{eq:asy_sol_field}
\end{equation}
The  on-shell action reduces to the surface term on the AdS boundary. Substituting the asymptotic form of the scalar field, Eq. \eqref{eq:asy_sol_field} into Eq. \eqref{eq:onshell_quase} yields 
\begin{equation}\label{eq:responsegeneralgkpw}
	\left \langle \mathcal{O} \right \rangle_S = 4r_0^4 \upvarphi^{(1)} \upvarphi^{(0)} = \updelta \left \langle \mathcal{O}\right\rangle.
\end{equation}
Relating this result to Eq. \eqref{eq:response_def} determines the retarded Green's function,
\begin{equation}
 G_R^{\mathcal{O}, \mathcal{O}} (\bm{\mathfrak{q}} = \bm{0}) = -4 r_0^4 \upvarphi^{(1)}.
\end{equation}
\noindent 

\section{The AdS$_5$--Schwarzschild deformed black brane}
\label{sec:eta_s_sads_5}

\par The general solution to 5D vacuum Einstein gravity with a negative cosmological constant depends on the horizon metric $H_{ij}$ and an integration constant, $k$. Provided that the constraint $R_{ij}=3kH_{ij}$ holds, the solution for $k=0$, leading to a planar horizon i.e. $H_{ij}=\delta_{ij}$, is the AdS$_5$--Schwarzschild black brane \cite{Aminneborg:1996iz}. The dual theory is a conformal fluid \cite{Bilic:2014dda}. Hence its stress-energy tensor is traceless, fixing the bulk viscosity \cite{rangamani, hubeny_hard}, $\zeta = 0$, leaving the shear viscosity $\eta$ as the only non-trivial transport coefficient  \cite{son_hydro, Son:2009zzc}. 
We will present the arguments and a similar calculation, when considering the deformed AdS$_5$--Schwarzschild  black brane as the gravitational background. The saturation of the $\frac{\eta}{s}$ ratio in the AdS$_5$--Schwarzschild black brane gravitational background reads \cite{kss}
\begin{equation}
 \frac{\eta}{s} = \frac{1}{4\pi}.
 \label{eq:eta_s_sads_5}
\end{equation}
One does not need discuss specific bulk features, as the existence of  solutions to the higher-dimensional Einstein's equations describing gravity is undertaken by
the Campbell--Magaard embedding theorems \cite{Bronnikov:2003gx}. 

\par \textcolor{black}{There is a correspondence between AdS/CFT and braneworld scenarios.} 
In an AdS bulk with cosmological constant $\Uplambda$, a solution must satisfy the effective Einstein's equations 
\begin{eqnarray}
R_{AB}=\Uplambda\,g_{AB}+\mathcal{E}_{AB}
\,,
\label{D+1eq}
\end{eqnarray}
\bltx{where $A,B=0,1,2,3,5,6$}. 
One can project Eq. \eqref{D+1eq} onto a timelike, \bltx{codimension-1, embedding AdS manifold}, in Gaussian 
coordinates $x^M$ = \bltx{($x^\mu, x^5$) -- for $\mu=0,1,2,3$, where $x^5=r$.} When $r=0$, it
corresponds to the brane itself, requires the Gauss--Codazzi equations to represent the embedding bulk Ricci tensor, when the discontinuity of the extrinsic curvature is related to the embedding codimension-1 bulk stress-tensor\footnote{This model emulates the one in Sect. 10.3 of Ref. \cite{maartens}.}.
 Hence, the field equations yield the effective Einstein's field equations on the bulk, whose corrections consist of an AdS bulk Weyl fluid \cite{ssm1}. This fluid flow is implemented by the bulk Weyl tensor, whose projection, the so called electric part of the Weyl tensor, reads  
\begin{eqnarray}
\!\!\!\!\!\!\!\!\mathcal{E}_{MN}(\sigma^{-1}) \!&=&\!-\frac{6}{\sigma}\!\left[ \mathcal{U}\!\left(\!u_M u_N \!+\! \frac{1}{3}h_{MN}\!\right) \!\right.\nonumber\\&&\left.\qquad\quad+ \mathit{Q}_{(M} u_{N)}\!+\!\mathcal{P}_{MN}\right], \label{A4}
\end{eqnarray}
\noindent \bltx{for $M,N=0,1,2,3,5$}, where $h_{MN}$ denotes the projector operator that is orthogonal to the  velocity, $u^M$, associated to the  Weyl fluid flow. In addition, $\mathcal{U}=-\frac16\sigma\mathcal{E}_{MN} u^M u^N$ is the effective energy density; $\mathcal{P}_{MN}=-\frac16\sigma\left(h_{(M}^{\;P}h_{N)}^{\;Q}-\frac13 h^{PQ}h_{MN}\right)\mathcal{E}_{PQ}$ is the  effective non-local anisotropic stress-tensor; and the effective non-local energy flux, $\mathit{Q}_M = -\frac16\sigma h^{\;P}_{\mu}\mathcal{E}_{PN}u^M$, is originated from the bulk free gravitational field. The tension is described by $\sigma$. Local corrections are encoded into the tensor \cite{ssm1,Shiromizu:2001jm}:	
\begin{eqnarray}\label{smunu}
\!\!\!S_{MN} \!=\! \frac{T}{3}T_{MN}\!-\!T_{MP}T^P_{\ N} \!+\! \frac{g_{MN}}{6} \Big[3T_{PQ}T^{PQ} - T^2\Big]
\end{eqnarray}
\noindent where $T_{MN}$ is the matter stress-tensor \bltx{and $T=T^{M}_{\;M}$ denotes the trace of $T_{MN}$}. 
The trace of $S_{MN}$ corresponds to the trace anomaly of the cutoff CFT on the brane \cite{maartens}. 
Higher-order terms in Eq. (\ref{smunu}) are neglected, as the embedding bulk matter density
is negligible.  Denoting by $G_{MN}$ the Einstein tensor, the 5D  Einstein's effective field equations read
\begin{equation}
G_{MN}
=T_{MN}+\mathcal{E}_{MN}(\sigma^{-1})+\bltx{\frac{1}{4\sigma}S_{MN}}=0 . \label{projeinstein}
\end{equation} 
Since $\mathcal{E}_{MN} \sim \sigma^{-1}$, it is straightforward to notice that in the infinitely rigid  limit,  $\sigma \rightarrow \infty$, GR is recovered and the Einstein's field equations have the standard form  $G_{MN} = T_{MN}$. 
Alternatively, the system of equations below is weaker than the effective field equations, and can be seen as  constraints
\begin{eqnarray}
R_{M w}=0,\ \ \
\mathring{R}=\Lambda,
\label{Deq}
\end{eqnarray} \bltx{where $\bltx{w=x^6}$ is the bulk extra dimension; $\mathring{R}$ and $\Lambda$ denote, respectively, the codimension-1 embedding bulk Ricci scalar and the 5D cosmological constant.} Eqs.~(\ref{Deq}) mimic constraints in the ADM procedure \cite{adm}, whereas the equation
$R_{MN}=\mathcal{E}_{MN}$ completes this system.

\bltx{One supposes a general metric,  setting the AdS radius to unity, 
\begin{eqnarray}
\!\!\!\!\!\!\!\!\!\!ds^2 = -{r^2} N(r) \mathrm{d}t^2 + \frac{1}{r^2 A(r)} \mathrm{d}r^2 + {r^2} \updelta_{ij} \mathrm{d}x^i \mathrm{d}x^j. \label{1a}
\end{eqnarray}}
\noindent 
\!By demanding that the ADM constraint leads to the AdS$_5$--Schwarzschild metric 
when $\beta\to1$, \bltx{and denoting by a prime the derivative with respect to $r$,} the Hamiltonian constraint reads,
\begin{eqnarray}\label{ctr}
&&{2N''(r)\over N(r)}-
{N'^2(r)\over N^2(r)}+{2A''(r)\over A(r)}
+{A'^2(r)\over A^2(r)}
-{N'(r)A'(r)\over N(r)A(r)}
\nonumber \\
&&\quad +{4\over r}\,\left({N'(r)\over N(r)}-{A'(r)\over A(r)}\right)
-{4A(r)\over r^2}=f(r,r_0,\beta),
\end{eqnarray}
where the function $f(r,r_0,\beta)$ is given by Eq. (\ref{r10}) in the Appendix \ref{app}.

\bltx{In the $u$ variable, 
the metric (\ref{1a}) reads}
\begin{eqnarray}
\!\!\!\!\!\!\!\!\!ds^2 = -\frac{r_0^2}{u^2} N(u) \mathrm{d}t^2 + \frac{1}{u^2 A(u)} \mathrm{d}u^2 + \frac{r_0^2}{u^2} \updelta_{ij} \mathrm{d}x^i \mathrm{d}x^j, \label{1}
\end{eqnarray}
The constraint (\ref{ctr}) is satisfied by 
\begin{eqnarray}
N(u) &=& 1 - u^4 + \left (\beta - 1 \right ) u^6,\label{eq:Nu}\\
A(u) &=& \left (1 - u^4 \right ) \left ( \frac{2 - 3u^4}{2- \left (4\beta-1\right ) u^4}\right ).
\label{eq:Au}
\end{eqnarray}
The constant $\beta$ parameter is referred to as a deformation parameter. \textcolor{black}{In the next section we will investigate how the shear-viscosity-to-entropy density ratio can drive specific values for $\beta$.}

\subsection{\textcolor{black}{Thermodynamics}}
\label{termo}
\par \textcolor{black}{Combining the metric \eqref{1}, with coefficients (\ref{eq:Nu}, \ref{eq:Au}), and the GKPW relation \cite{Witten:1998qj, gkp2}, we are able to obtain the partition function associated to the dual theory, and calculate the thermodynamic functions such as entropy, pressure and free energy. Basically,  the following action must be evaluated}\textcolor{black}{
\begin{align}
\begin{aligned}	\label{Sonshell1}
S_{E}\!=\!-\frac{1}{16\pi G}&\!\overbrace{\int \!d^{5}x\sqrt{g}\left(R\!-2\!\Lambda_5\right)}^{I_{\rm bulk}}\! \\ 
& -\!\frac{1}{8\pi G}\overbrace{\lim_{u\to 0}\int \!d^{4}x\sqrt{h}K}^{I_{GH}}\!+\!I_{\text{c.t}}\ ,	
\end{aligned}
\end{align}				}
\textcolor{black}{ where the first term is the Einstein--Hilbert action with the cosmological constant, the second term is the Gibbons--Hawking term, and the last is the counter term, which is introduced to ensure that the result is finite. In this case one uses the Euclidean signature, obtained by performing a Wick rotation in the time coordinate $t\mapsto i\tau$. This implies that $\tau$ is a periodic coordinate with period $2\pi$ \cite{Wald:1995yp}}.

Each term will be individually computed, starting by the Einstein--Hilbert term. The cosmological constant is $-2\Lambda_5=12$, and the expansion on $u$ of the scalar curvature reads 
\begin{eqnarray} \label{scalCurv}
R=-20-8\left(\beta-1\right)u^{4}+\ldots\ ,
\end{eqnarray}
since the variable $u$ is defined from $0$ to $1$. For the metric determinant, the expansion on $u$ is given by 
\bltx{\beq
	\!\!\sqrt{g}&\approx&\frac{r_0^4}{u^5}-\frac{\left(\beta-1\right)r_{0}^{4}}{u}+\frac{1}{2}(\beta-1)r_{0}^{4}u\nonumber\\&&+\frac{\left(1-\beta\right)}{4}\left[6-\left(1-\beta\right)\right]r_{0}^{4}u^{3}.\label{281}
\eeq}
Hence, the Einstein--Hilbert term becomes
\bltx{\begin{align} \label{Ibulk}
\begin{aligned}
\!\!\!\!I_{\text{bulk}}&=\left[\left(\frac{1}{\epsilon^{4}}-1\right)-2(\beta-1)\right.\\&\left.+\frac{1}{2}\left(\beta^{2}+2(\beta-1)^{2}+\beta-2\right)\right]\!,
\end{aligned}
\end{align}}
where $\epsilon\to 0$ is used to keep track of divergent terms, which will be cancelled with the counter term.

\par \textcolor{black}{The Gibbons--Hawking term is a surface term. 
By considering the normal vector $n_{\alpha}=g_{uu}^{-1/2}\delta_{\alpha}^{u}$, the induced metric for a hypersurface at constant $u$ is given by $h_{\mu\nu}=g_{\mu\nu}-n_{\mu}n_{\nu}$, using $g_{\mu\nu}$ from \eqref{1} we have
\begin{eqnarray}  \label{inducmetric}
ds_{\text{HS}}^{2}=-\frac{r_{0}^{2}}{u^{2}}N(u)dt^{2}+\frac{r_{0}^{2}}{u^{2}}\delta_{ij}dx^{i}dx^{j}.
\end{eqnarray}
The computation of $K$ is straightforward, being its expansion near the boundary given by
\begin{eqnarray} 
K=-4\left[1+\left(\beta-1\right)u^{4}+\ldots\right]\ ,
\end{eqnarray}
as well as for the metric determinant
\begin{eqnarray} 
\sqrt{h}=r_{0}^{4}\left[\frac{1}{u^{4}}-\frac{1}{2}+\frac{u^{2}}{2}\left(\beta-1\right)-\frac{u^{4}}{8}+\cdots\right]\ . \label{hdet}
\end{eqnarray}
Then, it is just a matter of manipulating terms to find
\bltx{\begin{eqnarray}
I_{GH}=-4\left[\frac{1}{\epsilon^{4}}-\frac{1}{2}\left(3-2\beta\right)\right]\ ,
\end{eqnarray}}
where again, the divergent term is left explicit.}

\par \textcolor{black}{In dimension $d$, the counter term has a standard form and depends only on the geometry of the boundary theory, explicitly given by  \cite{Emparan:1999pm}
\begin{align}
\begin{aligned}
	\!\!\!\!I_{\text{c.t}}=&\frac{1}{8\pi G}\!\lim_{u\to 0}\int d^{d}x\sqrt{h}\left\{ \left(d-1\right)+\frac{\mathfrak{R}}{2\left(d-2\right)}\right.\\
	&\left.\!\!\!\!\!\!+\frac{1}{2\left(d\!-\!4\right)\!\left(d\!-\!2\right)^{2}}\!\left[\mathfrak{R}_{\mu\nu}\mathfrak{R}^{\mu\nu}\!-\!\frac{d\,\mathfrak{R}^{2}}{4\!\left(d\!-\!1\right)}\right]+\ldots\right\} 
\end{aligned}
\end{align}
where $\mathfrak{R}$ and $\mathfrak{R}_{\mu\nu}$, respectively, refer to the scalar curvature and Ricci tensor of the induced metric \eqref{inducmetric}, (remembering that ${\footnotesize{\mu, \nu = 0,1,2,3}}$), and one can quickly check that these vanish. In dimension $d=4$, remembering that it is a surface term, it leads to the following, 
\begin{eqnarray} 
I_{\text{c.t}}=\frac{3}{8\pi G}\lim_{u\to 0}\int d^{4}x\sqrt{h}\ .
\end{eqnarray} 
Eq. \eqref{hdet} yields 
\begin{eqnarray}
I_{\text{c.t}}=\frac{3r_{0}^{4} V b}{\bltx{8\pi G}}\left[\frac{1}{\epsilon^{4}}-\frac{1}{2}\right],
\end{eqnarray}
where $V=\int dxdydz$ and $b=\int d\tau$. (Usually this is called $\beta$ in the literature, but to avoid confusion with the deformation parameter, we called it $b$.) Combining the integrals and restoring the constant factors yields 
\begin{equation} \label{partfnc}
\!\!\!\!\bltx{S_{E}=\frac{Vbr_{0}^{4}}{8\pi G}\left( \frac{11-15\beta+3\beta^{2}}{2}\right).}
\end{equation}
 Eq.  \eqref{partfnc} is the partition function of the dual theory at the boundary, according to the GKPW relation. Now, from statistical mechanics one knows that $Z=bF$, where $F$ is the free energy. Therefore we can calculate thermodynamic functions, by taking derivatives of $F$.}

\par \textcolor{black}{Since we are going to compute thermodynamic functions, it is convenient to know the temperature. In the AdS/CFT context, the temperature is associated to the Hawking temperature at the horizon of the black hole \cite{Liu:2014dva}
	\begin{eqnarray} \label{temperaturegeneral}
	T=\frac{1}{4\pi}\bltx{\lim_{u\to1}}\sqrt{\frac{\dot{g}_{tt}(u)}{\dot{g}_{rr}(u)}}.
	\end{eqnarray}
For the metric \eqref{1}, this expression is simply
	\begin{eqnarray} \label{Tofr}
	T=\frac{r_{0}}{\pi}\sqrt{\frac{\beta-2}{3-4\beta}}.
	\end{eqnarray}
	{\bltx{It is important to mention that expression \eqref{Tofr} is obtained by approximating the metric coefficients near the horizon, i.e. $g_{tt}(u=1)\approx g^{(0)}_{tt}(u=1)+g^{(1)}_{tt}(u=1)(u-1)+\ldots$, and similarly for $g_{uu}$.} Fig. \ref{fig:1} illustrates Eq. (\ref{Tofr}) as a function of $\beta$.}
\begin{figure}[H]
\centering\includegraphics[width=6.6cm]{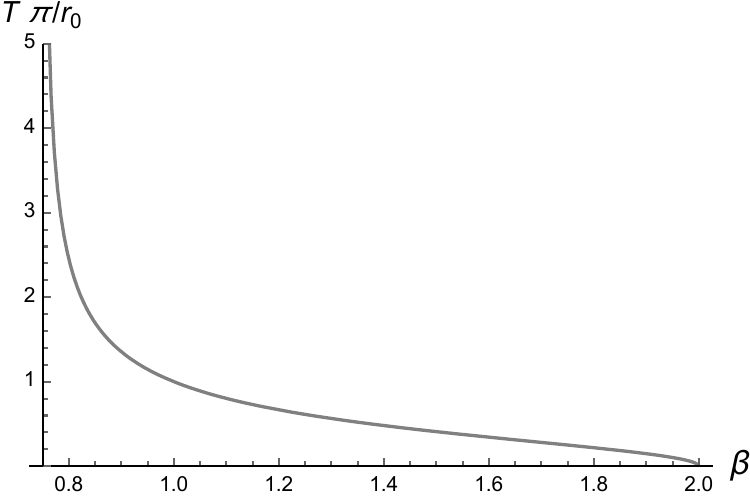}
\caption{Temperature of the deformed black brane, as a function of $\beta$.}
\label{fig:1}
\end{figure} 
\noindent {The deformed black brane temperature diverges at $\beta\to 3/4$, having imaginary values for either $\beta < 3/4$ or $\beta > 2$. As the deformed black brane temperature cannot attain divergent values or imaginary ones, the analysis of the deformed black brane temperature constrains the $\beta$ parameter in the open range $\beta\in(3/4,2)$. }
One can invert Eq. (\ref{Tofr}) to express $r_0$ as
	\begin{eqnarray} \label{rofT}
	r_{0}=\pi\sqrt{\frac{3-4\beta}{\beta-2}}T\ .
	\end{eqnarray} 
Finally, the free energy can be read off, when Eq. \eqref{rofT} is replaced into \eqref{partfnc}, yielding 	\bltx{\begin{align}
	\label{freenergyofT}
	\!\!\!\!\!\!\!F=\frac{\pi^{3}V}{8G}\left(\frac{11-15\beta+3\beta^{2}}{2}\right)\left(\frac{3-4\beta}{\beta-2}\right)^{2}T^{4}\ .
	\end{align}}
}

\par \textcolor{black}{The state functions can now be computed using standard statistical mechanics in the canonical ensemble}
\bltx{	\begin{eqnarray} \label{entropyofT}
			\!\!\!\!\!\!\!\!\!\!\!\!\!\!s&=&-\frac{1}{V}\frac{\partial F}{\partial T}\!=\!-\frac{\pi^{3}}{2G}\!\left(\frac{11\!-\!15\beta\!+\!3\beta^{2}}{2}\right)\!\!\left(\frac{3-4\beta}{\beta-2}\right)^{\!\!2}\!T^{3},\\
			\!\!\!\!\!\!\!\!P&\!=\!&\!-\frac{\partial F}{\partial V}\!=\!-\frac{\pi^{3}}{8G}\!\left(\frac{11\!-\!15\beta\!+\!3\beta^{2}}{2}\right)\left(\frac{3\!-\!4\beta}{\beta-2}\right)^{2}\!T^{4}\ ,\label{pofT}\\
			\!\!\!\!\!\!\varepsilon &\!=\!& \frac{F}{V}\!-\!Ts \!=\!\frac{5\pi^{3}}{8G}\!\left(\frac{11\!-\!15\beta\!+\!3\beta^{2}}{2}\right)\!\left(\frac{3-4\beta}{\beta-2}\right)^{2}\!T^{4}\label{espsofT}.
	\end{eqnarray}}
\bltx{Despite the negative sign in front of entropy and pressure, these quantities are positive in the range of $\beta$ to be considered in the analysis to come in the next session.}
{For a perfect fluid, the energy-momentum tensor reads
\begin{equation}
T^{ab}=\left(\varepsilon+P\right)u^{a}u^{b}+Pg^{ab}.\label{emt}
\end{equation}
From Eqs. (\ref{pofT}, \ref{espsofT}), evaluated at the
boundary,  the trace of the energy-momentum tensor (\ref{emt})  
is given by 
\bltx{\beq
\!\!\!\!\!\!\!\!\! g_{\mu\nu}T^{\mu\nu}&\!=\!&-\varepsilon+3P\nonumber\\
 & =&-\frac{\pi^{3}}{G}\left(\frac{11-15\beta+3\beta^{2}}{2}\right)\!\left(\frac{3-4\beta}{\beta-2}\right)^{\!\!2}T^{4}.\label{trEMT}
\eeq}
}

For future reference, changing $T$ to $r_0$ using \eqref{rofT}, the entropy density in Eq. \eqref{entropyofT} can be written as, 
\bltx{\begin{eqnarray} \label{sofr}
s=-\frac{r_{0}^{3}}{2G}\left(\frac{11-15\beta+3\beta^{2}}{2}\right)\left(\frac{3-4\beta}{\beta-2}\right)^{1/2} \,.
\end{eqnarray}}
As the entropy of a black hole obtained from Einstein's equations is proportional to its area, in the particular case of metric \eqref{1} we have a deformation of a Schwarzschild black hole that is asymptotically AdS. This deformation breaks the spherical symmetry of our problem, and we have just used the AdS/CFT correspondence to compute the surface area of the black hole, i.e. $A=4GsV$.

\section{\texorpdfstring{$\frac{\eta}{s}$}{} for the AdS$_5$--Schwarzschild deformed black brane}
\label{sec:eta_s_ads_mgd_5}

As metric \eqref{1} arises from a deformation of the AdS$_5$--Schwarzschild \cite{mgd1}, the same action-dependent results may be applied.  The metric determinant,  $g$, is such that $
\sqrt{-g} = \frac{r_0^4}{u^5} \sqrt{\frac{N}{A}}$, where, from now on, $N$ and $A$  refer respectively to $N(u)$ and $A(u)$.

Consider a bulk perturbation $h_{xy}$ such that:
\begin{equation}\label{ref12}
 ds^2 = ds^2_{{\rm AdS}_5-{SD}} + 2h_{xy} \mathrm{d}x \mathrm{d}y \ ,
\end{equation}
\noindent where $ds^2_{{\rm AdS}_5-{SD}}$ denotes the AdS$_5$--Schwarzschild deformed  black brane metric, Eq. \eqref{1}. \textcolor{black}{In appendix \ref{appB} we show that the field associated to the perturbation propagates with the speed of light, this signals that no anomaly is present when it comes to the spacetime causal structure. }

Recall  Eq. \eqref{oooo}, for $h_{xy}^{(0)}$ being the perturbation added to the boundary theory, which is asymptotically related to $h_{xy}$, the bulk perturbation, by\footnote{We are now using the $u$ coordinate, instead of $r$.}
\begin{equation}
 g^{xx}h_{xy} \sim h_{xy}^{(0)} \left ( 1 + h_{xy}^{(1)} u^4 \right ) \ ,
 \label{eq:perturb_asym_def}
\end{equation}
\noindent according to Eq. \eqref{eq:asy_sol_field}. Notice that one can directly use the results for a massless scalar field, as  $g^{xx}h_{xy}$ obeys the EOM for a massless scalar field \cite{Son:2009zzc, son_hydro}. Besides, the deformed AdS$_5$--Schwarzschild  black brane has the same asymptotic behavior of the AdS$_5$--Schwarzschild black brane (namely, Eq. \eqref{eq:asym_ads}). One can identify $g^{xx}h_{xy}$ as the bulk field, $\upvarphi$, which plays the role of an external source of a boundary operator, in this case $\tau^{xy}$. Therefore, one can directly obtain the response $\updelta \left \langle \tau^{xy} \right \rangle$, from Eq. \eqref{eq:responsegeneralgkpw}, 
\begin{equation}
\updelta \left \langle \tau^{xy} \right \rangle = \frac{r_0^4}{16 \pi G}4 h_{xy}^{(1)}h_{xy}^{(0)} \ , 
\label{eq:response_2}
\end{equation}
\noindent where it is now convenient to reintroduce the $1/16 \pi G$ factor. Comparing Eqs. \eqref{oooo} and \eqref{eq:response_2} yields 
\begin{equation} \label{eq:eta_quase}
 i \omega \eta = \frac{r_0^4}{4\pi G} h_{xy}^{(1)} .
\end{equation}
Taking the ratio between Eq. \eqref{eq:eta_quase} and the entropy \eqref{sofr} we find
\bltx{\begin{eqnarray}
\!\!\!\!\!\!\!\!\!\!\!\!\!\!\frac{\eta}{s}=-\frac{r_{0}}{\pi}\left[\left(\frac{1}{11-15\beta+3\beta^{2}}\right)\left(\frac{\beta-2}{3-4\beta}\right)^{1/2}\right]\frac{h_{xy}^{\left(1\right)}}{i\omega},
\label{eq:eta_s_geral}
\end{eqnarray}}
\noindent \!\!where $h_{xy}^{(1)}$ is the solution of the EOM for the perturbation $g^{xx}h_{xy} \equiv \upvarphi$, which is that of a massless scalar field \cite{Son:2009zzc, son_hydro}
\begin{equation}
	\nabla_M \left( \sqrt{-g} g^{MN} \nabla_N \upvarphi \right) = 0\ .
\end{equation}
Considering a stationary perturbation, given by the form $\upvarphi(u,t) = \upphi(u) e^{-i\omega t}$,  the perturbation equation reduces to a second-order ODE for $\upphi(u)$,
\begin{equation}
\ddot\upphi + \frac12\left ( \frac{\dot{N}A}{2 N}+\frac{N\dot{A}}{A} - \frac{3}{u}\right ) \dot\upphi + \frac{1}{NA} \frac{\omega^2}{r_0^2} \upphi = 0 \ .
\label{eq:pertu_edo2}
\end{equation}
To derive the solution of Eq. \eqref{eq:pertu_edo2}, two boundary conditions are imposed: the incoming wave boundary condition in the near-horizon region, corresponding to $u \rightarrow 1$, and a Dirichlet boundary condition at the AdS boundary, $\upphi (u\rightarrow 0) = \upphi^{(0)}$, where $h_{xy}^{(0)} =  \upphi^{(0)} e^{-i \omega t}$. 

\par The incoming wave boundary condition near the horizon is obtained by solving Eq. \eqref{eq:pertu_edo2} in the limit $u \rightarrow 1$. After a straightforward computation one finds the following
\begin{eqnarray}
 \upphi \propto \exp \left (\pm i \frac{\omega}{r_0} \sqrt{\frac{4\beta -3}{\beta - 1}}\sqrt{1-u}\right).
\end{eqnarray}
This solution has a natural interpretation using tortoise coordinates, allowing one to identify it as a plane wave \cite{natsuume}. The positive exponent represents an outgoing wave, whereas the negative one describes the wave incoming to the horizon, which, according to the near-horizon boundary condition, allows us to fix
\begin{equation}
\upphi \approx \exp \left (- i \frac{\omega}{r_0} \sqrt{\frac{4\beta -3}{\beta - 1}}\sqrt{1-u}\right).
\label{eq:sol_nh}
\end{equation}

\par Next we solve Eq. \eqref{eq:pertu_edo2} for all $u\in[0,1]$ as a power series in $\omega$. As we are interested in the hydrodynamic limit of this solution, i.e. $\omega \rightarrow 0$, it is sufficient to keep the series up to linear order:
\begin{equation}
 \upphi(u) = \Phi_0(u) + \omega \Phi_1(u) \ .
 \label{eq:sol_omega_power}
\end{equation}
Since the second term in Eq. \eqref{eq:pertu_edo2} is of order $\omega^2$, it can be neglected. By direct integration the solution reads
\begin{equation}
	\Phi_i = C_i + K_i \int \frac{u^3}{\sqrt{N(u)A(u)}}\, \mathrm{d} u\ , 
\end{equation}
for $C_i$ and $K_i$ the integration constants and $i=0,1$. Thus, according to Eq. \eqref{eq:sol_omega_power}, we have 
\begin{equation}
 \upphi = \left (C_0 + \omega C_1 \right ) + \left ( K_0 + \omega K_1 \right ) \int \frac{u^3}{\sqrt{N(u)A(u)}}\, \mathrm{d} u \ .
 \label{eq:sol_omega_power_general}
\end{equation}
In order to impose the boundary conditions we expand the integral (\ref{eq:sol_omega_power_general}) around $u\rightarrow 0$ and $u\rightarrow 1$. It yields, up to leading order in the respective expansions,
\begin{eqnarray}
\!\!\!\!\!\!\!\!\!\!\!\!\!\!\int\! \frac{u^3}{\sqrt{NA}}\, \mathrm{d} u  \!&=&\!
\begin{cases} \frac{u^4}{4} \ ,&\;\;\;\text{for}\;\;u \rightarrow 0,\\
\frac{3-4\beta}{\beta-1}\sqrt{\frac{\beta-1}{3-4\beta}}\sqrt{1-u} \ ,&\;\;\;\text{for}\;\;u \rightarrow 1.\end{cases}
\end{eqnarray}
The first pair of integration constants is fixed by the Dirichlet boundary condition
\begin{eqnarray}
\!\!\!\!\!\!\!\!\!\!\!\!\!\! \lim_{u\rightarrow0} \upphi =\left (C_0 + \omega C_1 \right ) + \left ( K_0 + \omega K_1 \right ) \lim_{u\rightarrow0} \frac{u^4}{4} =  \upphi^{(0)},
\end{eqnarray} 
implying that $\left (C_0 + \omega C_1 \right ) = \upphi^{(0)}$. Near the horizon one has 
\begin{equation}
   \upphi \approx \upphi^{(0)} - \left ( K_0 + \omega K_1 \right ) \frac{(4\beta-3)}{\beta-1}\sqrt{\frac{\beta-1}{4\beta-3}}\sqrt{1-u}.
   \label{eq:sol_general_nh}
\end{equation}
Expanding Eq. \eqref{eq:sol_nh} up to $\mathcal{O}(\omega)$ yields
\begin{equation}
   \upphi \propto 1 - i \frac{\omega}{r_0}\sqrt{\frac{4\beta-3}{\beta-1}}\sqrt{1-u}.
\end{equation}
It is straightforward to see that Eq. \eqref{eq:sol_general_nh} fixes the proportionality according to
\begin{equation}
   \upphi \approx \upphi^{(0)} - i \upphi^{(0)}\frac{\omega}{r_0}\sqrt{\frac{4\beta-3}{\beta-1}}\sqrt{1-u}.
   \label{eq:sol_nh_power_omega}
\end{equation}
Comparison between Eqs.\eqref{eq:sol_general_nh} and \eqref{eq:sol_nh_power_omega} immediately fixes the second pair of integration constants:
\begin{eqnarray}
\left ( K_0 + \omega K_1 \right ) = i \upphi^{(0)}\frac{\omega}{r_0} \left (\frac{\beta -1}{4\beta-3} \right ) \frac{|4\beta - 3|}{|\beta-1|}.
\end{eqnarray}
\noindent Then the full solution reads 
\begin{equation}
 \!\!\!\upphi \!=\! \upphi^{(0)} \left ( 1 \!+\! i \frac{\omega}{r_0} \left (\frac{\beta \!-\!1}{4\beta\!-\!3} \right ) \frac{|4\beta \!-\! 3|}{|\beta\!-\!1|} \int \!\frac{u^3}{\sqrt{NA}}\, \mathrm{d} u \right ) \ . 
\end{equation}

\textcolor{black}{Accordingly, the full time-dependent perturbation }
\begin{equation}
	\upvarphi = g^{xx}h_{xy} = \upphi(u) e^{-i\omega t}\ ,
\end{equation}
\textcolor{black}{is asymptotically given by:}
\begin{equation}
g^{xx}h_{xy} \sim  e^{-i\omega t}\upphi^{(0)} \left ( 1 + i \frac{\omega}{r_0} \left (\frac{\beta -1}{4\beta-3} \right ) \frac{|4\beta - 3|}{|\beta-1|} \frac{u^4}{4}\right ) \ .
\label{eq:perturb_asym_solution}
\end{equation}
\textcolor{black}{Eqs. (\ref{eq:perturb_asym_def}, \ref{eq:perturb_asym_solution}) yield 
\begin{equation}
h_{xy}^{(1)} = \frac{i\omega}{4r_0} \left (\frac{\beta -1}{4\beta-3} \right ) \frac{|4\beta - 3|}{|\beta-1|} \ ,
\label{eq:key}
\end{equation}
where $h_{xy}^{(0)} =  \upphi^{(0)} e^{-i \omega t}$.} \textcolor{black}{The term multiplying $\frac{i\omega}{4r_0}$ in Eq. \eqref{eq:key} can be visualized in the following plot:}
\begin{figure}[H]
\centering\includegraphics[width=6.6cm]{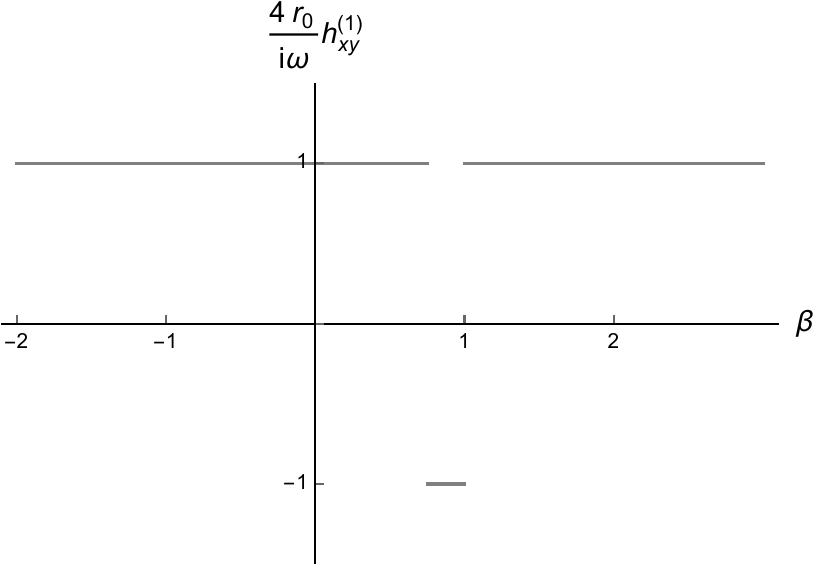}
 \caption{$4r_0 \frac{h_{xy}^{(1)}}{i\omega}$ as a function of $\beta$.}
\label{fig}
\end{figure}
Therefore we have different signs depending on the value of $\beta$, 
\begin{equation}\label{ank}
\begin{cases} 
h_{xy}^{(1)}=-\frac{i\omega}{4r_{0}} & \frac{3}{4}<\beta<1 \ , \\
h_{xy}^{(1)}=\frac{i\omega}{4r_{0}} & \beta<\frac{3}{4}\text{ or }\beta>1 \ . 
\end{cases} 
\end{equation}
A negative value for $h_{xy}^{(1)}$, without further constraints, would imply a negative value of $\frac{\eta}{s}$, i.e., a negative viscosity or entropy density, which would violate the second law of thermodynamics. Therefore, demanding thermodynamical consistency leads to the following first bound in the deformation parameter: either $\beta < \frac{3}{4} \ \text{or} \ \beta > 1 $.




\textcolor{black}{Now, substituting \eqref{eq:key}  in Eq. \eqref{eq:eta_s_geral} yields }
\bltx{\begin{equation} \label{eq:etaSfinal}
\frac{\eta}{s}=\begin{cases}
-\frac{1}{4\pi}\left(\frac{1}{11-15\beta+3\beta^{2}}\right)\left(\frac{\beta-2}{3-4\beta}\right)^{1/2}\!\!, & \beta>1\\
\frac{1}{4\pi}\left(\frac{1}{11-15\beta+3\beta^{2}}\right)\left(\frac{\beta-2}{3-4\beta}\right)^{1/2}\!\!, & \beta<1
\end{cases} \ .
\end{equation}	}

\begin{figure}[H]
\centering\includegraphics[width=7cm]{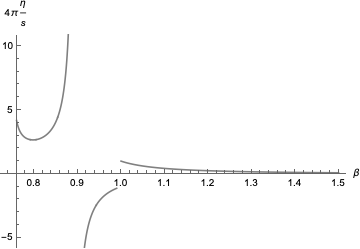}
\caption{$\frac{\eta}{s}$ ratio of the deformed black brane, as a function of $\beta$.}
\label{fi2}
\end{figure}

\textcolor{black}{Fig. \ref{fi2} illustrates Eq. (\ref{eq:etaSfinal}) as a function of $\beta$.}
\noindent \textcolor{black}{For the precise value $\beta =1$, the deformed black brane $\frac{\eta}{s}$ ratio is exactly $\frac{1}{4\pi}$, recovering the KSS result for the AdS$_5$--Schwarzschild   black brane. Besides, Fig. \ref{fi2} shows the divergence of $\frac{\eta}{s}$ for $\beta \approxeq 0.9$ as well as the vanishing of the $\frac{\eta}{s}$ ratio, for $\beta = 2$.}

\textcolor{black}{Therefore, \pptx{\emph{a priori}} the deformation parameter can attain the \bltx{ranges 
\begin{equation}\label{ra1}
0.75< \beta < 0.9\;\;\;\text{and}\quad 1 < \beta \leq 2.
\end{equation}
The value $\beta \leq 2$ is seen from \eqref{eq:etaSfinal}, since $\beta=2$ makes that quantity equal to zero, whereas the range  $0.9\leq\beta < 1$ imply $\frac{\eta}{s}<0$, which has no physical significance.} The saturation $\frac{\eta}{s} = \frac{1}{4\pi}$, corresponding to  to the infinite 't Hooft coupling limit \cite{Cremonini:2011iq}}, then implies $\beta = 1$. This result  has been expected, as this case recovers the AdS$_5$--Schwarzschild black brane (\ref{eq:sads_5_u}). \pptx{However, an additional consistence test must take into account Eq. (\ref{eq:Nu}), that defines the deformed AdS$_5$--Schwarzschild black brane event horizon. In fact, let us call by $u_\beta = 1/r_\beta$ the solution of the algebraic equation $N(u)=0$, in (\ref{eq:Nu}). The first consistence test must regard the choice of $\beta$ in such a way that it produces a real event horizon\footnote{Equivalently, that the algebraic equation $N(u)=0$, in (\ref{eq:Nu}) does not have only complex solutions.}. Therefore, this restricts more the possible range for $\beta$, from $1 < \beta \leq 2$ to $1 < \beta \leq 1.384$. A second consistence test involves the fact that the  $r_0=\lim_{\beta\to1} r_\beta$ horizon, corresponding to the standard AdS$_5$--Schwarzschild black brane event horizon,  is of Killing type. Along our previous calculations, the horizon is assumed to be at $r_0$. For it to be a good approximation in the proposed ranges of $\beta$, in such a way that $\left|r_0-r_\beta\right|\lesssim 10^{-2}$,
we must restrict a little more the allowed range to $1 < \beta \lesssim  1.2$, since for the another range $0.75< \beta < 0.9$ the condition $\left|r_0-r_\beta\right|\lesssim 10^{-2}$ already holds. Hence, the $\beta$ parameter is restricted into the ranges
\begin{equation}\label{ra2}
0.75< \beta < 0.9\;\;\;\text{and}\quad 1 < \beta \lesssim 1.2.
\end{equation}}

\textcolor{black}{To end this section we present a comparison between results obtained with metric \eqref{1} and the conventional AdS$_5-$Schwarzschild, which also gives us insight on the effect of the parameter $\beta$. Denoting $T_S$, $s_S$ and $\left(\frac{\eta}{s}\right)_{S}$ the temperature, entropy density and shear viscosity to entropy density of the standard AdS$_5-$Schwarzschild spacetime, respectively, one can check that the corresponding {positive} quantities for fixed \bltx{$\beta=1.05$} are}

\begin{align}
	\begin{aligned} \label{compar1}
	\textcolor{black}{\!\!\!\!\!T=0.89T_{S},\qquad s=1.82s_{S},\qquad \frac{\eta}{s}=0.54\left(\frac{\eta}{s}\right)_{S}},
	\end{aligned}
\end{align}
For instance, \textcolor{black}{if $\beta=1.2$ one finds}
\bltx{	\begin{align}
	\begin{aligned} \label{compar2}
	\!T=0.67T_{S},\qquad s=6.03s_{S},\qquad \frac{\eta}{s}=0.17\left(\frac{\eta}{s}\right)_{S}\ .
	\end{aligned}
	\end{align}}
	
Considering the results \eqref{compar1} and \eqref{compar2}, the effects of the deformation in the metric are clear,  changing thermodynamics and hydrodynamics by a numerical factor. \textcolor{black}{In the range $1<\beta\leq1.2$, there is a violation of the KSS bound. One can speculate that the violation comes from the fact that the solution under investigation does not obey Einstein's equations of GR, since it was obtained via an embedding in a higher dimensional space-time, whose evolution is governed by an equation that has the Einstein's field equations as a certain limit, c.f. Eq. \eqref{projeinstein}. }
\bltx{Fig. \ref{fi2} illustrates that the range $0.75<\beta < 0.9$ is formally allowed, wherein the deformation parameter makes the KSS bound not to be violated. The existence of a range where the KSS bound is violated, namely $1<\beta\lesssim 1.2$, but no pathologies in causality of space-time or thermodynamic functions can be seen, is also one of the main results of this work.} The meaning of the $\beta$ parameter will be further discussed in Sec. \ref{scru}. We emphasize that it is a free constant parameter, generating a family of deformed AdS$_5$--Schwarzschild black branes, which has been constrained for different reasons. We have imposed compliance with the second law of thermodynamics, thus discarding the ranges which would yield negative values of $\frac{\eta}{s}$. Therefore, the family of solutions obtained with the allowed values of $\beta$ can be an interesting result worthy further investigation, mainly in the AdS/QCD correspondence. \bltx{The embedding bulk scenario and ADM procedure, in which the deformed AdS${}_5$--Schwarzschild black brane was obtained, provides one more counterexample setup to the KSS bound conjecture. Besides, these results can play a relevant role on the QGP, whose measured viscosity is close to the
KSS bound, possibly violates the bound \cite{Cherman:2007fj}. In the next section we also address a possible scenario that corroborates to the violation of the KSS bound in the range $1<\beta\lesssim 1.2$.}

\section{Scrutinizing the $\beta$ parameter}
\label{scru}
{This section is devoted to clarify aspects of the $\beta$ parameter.
If one considers AdS/CFT in the braneworld, it relates the electric part of the Weyl tensor $\mathcal{E}_{\mu\nu}$ in Eq. (\ref{A4}), that represents (classical) gravitational waves in the bulk, to the expectation value $\langle T_{\mu\nu}\rangle$ of the (renormalized) energy-momentum tensor of conformal fields on the brane\footnote{The large $N$ limit expansion of the CFT requires $N\sim 1/(\sigma\ell_p)^2\gg1$. In the original Randall--Sundrum braneworld models, the Planck length, $\ell_p$ (for $8\pi G_4 = \ell_p^2$, where $G_4$ is the 4D Newton constant), is related to the 5D fundamental gravitational length $\ell_5$ by $\ell^2_p = \sigma \ell^3_5$ \cite{Randall:1999vf,ssm2}, where $\sigma$ is the brane tension.} \cite{Shiromizu:2001jm,Kanno:2002iaa}. Besides, the presence of the brane introduces a normalizable 4D graviton and an ultraviolet (UV) cut-off in the CFT, proportional to $\sigma^{-1}$. The general-relativistic limit requires $\sigma\to\infty$, corresponding to a geometric rigid  brane with infinite tension.  In the AdS/CFT setup, 
$\mathcal{E}^\nu_{\;\mu}\sim\ell_p^2 \langle T_{\;\mu}^{\nu}\rangle$. 
Since the electric part of the Weyl tensor is traceless, such a correspondence would imply that $\langle T_{\;\mu}^{\mu}\rangle\equiv \langle T \rangle=0$. In other words, it would hold in the case where the conformal
symmetry is not anomalous.  Eq. (\ref{trEMT}) therefore 
indicates a conformal anomaly due to the quantum corrections induced by $\beta$.
Eq. (\ref{trEMT}) yields $ \langle T \rangle\neq0$ for any value of $\beta$ but $\beta=1$. It is in full compliance with the fact that if $ \langle T \rangle=0$, then the UV cut-off would be required to be much shorter than any physical length scale involved.
Besides, $  \langle T \rangle=0$ for any value of $\beta$ would also demand the absence of any intrinsic
4D length associated with the background, otherwise
the CFT is affected by that scale.
For the deformed AdS$_5$--Schwarzschild black brane, the horizon radius  $r_0$ is a natural length scale and one therefore expects that only CFT modes with
wavelengths much shorter than $r_0$, that are much larger than $\sigma^{-1}$, can propagate freely.  Bulk perturbations at the boundary work as sources the CFT fields, and can produce  $\langle   T \rangle=0$.}

Of course, this requires that the UV cut-off be much shorter
than any physical length scale in the system.
For a black hole, the horizon  radius is a natural
length scale and one therefore expects that only CFT modes with
wavelengths much shorter than $r_{\rm h}$, that are still much larger than $\sigma^{-1}$, propagate
freely  \cite{Casadio:2003jc}.

Besides, for the deformed AdS$_5$--Schwarzschild black brane, one can emulate the holographic computation of the 
 Weyl anomaly \cite{Henningson:1998gx}. In fact, denoting $a$ and $c$ central charges of the conformal gauge theory, according to Eq. (24) of Ref. \cite{Cremonini:2011iq}, 
 \beq\label{wa}
\!\!\!\!\!\!\!\! \langle T^\mu_{\;\,\mu}\rangle_{\rm CFT}&=& \frac{c}{16\pi^2}\left(R_{\mu\nu\rho\sigma}R^{\mu\nu\rho\sigma}-2R_{\mu\nu}R^{\mu\nu}+\frac13 R^2\right)\nonumber\\
&& -\frac{a}{16\pi^2}\left(R_{\mu\nu\rho\sigma}R^{\mu\nu\rho\sigma}\!-\!4R_{\mu\nu}R^{\mu\nu}\!+\!R^2\right),
 \eeq where the terms in parentheses are, respectively, the Euler density and the square of the Weyl curvature. 
 
 \bltx{It is worth to mention the splitting of the allowed range of $\beta$ into $0.75<\beta<0.9$ and $1<\beta\lesssim 1.2$.
 Firstly, considering the range $1<\beta\lesssim 1.2$, Ref.  \cite{Kats:2007mq} studied an effective 5D bulk gravity dual, and showed that  the KSS bound is violated,  
whenever the central charges in the Weyl anomaly (\ref{wa}) satisfy $|c - a|/c \ll 1$. In this way, the inequality
$c>a$ yields the KSS bound to be violated \cite{Buchel:2008vz,viol1}. Ref. \cite{Kats:2007mq} showed that, as an effect of curvature squared corrections in the AdS bulk, the 
shear viscosity-to-entropy density ratio can be expressed as $\frac{\eta}{s}= \frac{1}{4\pi}\frac{a}{c}+\mathcal{O}(1/N^2)$.
Therefore, in the large $N$ limit, the equality $\frac{\eta}{s}\approxeq \frac{1}{4\pi}\frac{a}{c}$ holds, and the central charges ratio drive the KSS bound violation, whenever $c\neq a$. In fact, the well-known $\mathcal{N}=4$, SU($N$)  super-Yang--Mills theory implies $a = c$, however nothing precludes that $c\neq a$ in other cases \cite{Kats:2007mq}.} 

\bltx{Secondly, now considering the allowed range $0.75<\beta<0.9$,} the deformed AdS$_5$--Schwarzschild black brane, on the boundary $u\to0$, 
the square of the Weyl curvature can be expanded as
\beq\label{430}
\!\!\!\!\!\!\!N^2\left(\frac{40}{3}+\frac{32}{3} (\beta -1) u^4+8 (\beta -1) u^6\right)+\mathcal{O}\left(u^7\right),
\eeq
and  the Euler density as
\beq\label{120}
\!\!\!\!\!\!\!N^2\left(120+96 (\beta -1) u^4+72 (\beta -1) u^6\right)+\mathcal{O}\left(u^7\right),
\eeq where $N^2 = \pi L^3/2G$.
One notices in Eqs. (\ref{430}, \ref{120}) that the leading-order terms contain
factors $(\beta-1)u^p$, for $p=4,6$. Therefore, the limits $\beta\to 1$, corresponding to the standard AdS$_5$--Schwarzschild black brane, and the boundary $u\to0$ limit, are indistinguishable. Hence, the limit $u\to0$ yields 
\beq
 \langle T^\mu_{\;\,\mu}\rangle_{\rm CFT}=\frac{520 N^2}9,\eeq
having the same result of the standard AdS$_5$--Schwarzschild black brane.

It is worth to compare an already known result about $\frac{\eta}{s}$ in presence of quantum corrections. In fact, Ref. \cite{Myers:2008yi} discusses quantum corrections to the $\frac{\eta}{s}$ ratio, by including higher derivative terms with the 5-form RR flux to the calculation. Corrections are implemented as inverse powers of the colour number $N$, and the leading $1/N^2$ correction adds two corrections terms to entropy density, $s$, modifying $\frac{\eta}{s}$ in QCD strongly coupled QGP. Its original value, $\frac1{4\pi}$, is increased by approximately 37\%, roughly 22\% due to the first correction term and 15\% due to the second. As discussed in this section, \bltx{our setup yields corrections that can be interpreted as quantum ones, induced by $\beta$, as expressed in Eq. (\ref{trEMT}). For $\beta = 0.75$, consisting of a lower bound for $\beta$, the $\frac{\eta}{s}$ ratio increases $\sim4.1$ times the original $\frac{\eta}{s}=\frac1{4\pi}$ value. In the range $0.75<\beta < 0.9$, there is a minimum at $\beta\approx 0.8$, for which the shear viscosity-to-entropy ratio equals $2.5$ the KSS bound.
In the range $1< \beta \leq 1.2$, we showed that the KSS bound is violated. For example, as analyzed in Eq. (\ref{compar1}, \ref{compar2}), the value $\beta=1.05$ yields $\frac{\eta}{s}=0.54\left(\frac{\eta}{s}\right)_{S}$, whereas taking $\beta=1.2$ implies that $\frac{\eta}{s}=0.17\left(\frac{\eta}{s}\right)_{S}$.}

\section{Concluding remarks and perspectives} \label{sec:5}

The ADM procedure was used to derive a family of AdS$_5$--Schwarzschild deformed gravitational backgrounds, involving a free parameter, $\beta$, in the black brane metric  (\ref{1}, \ref{eq:Nu}, \ref{eq:Au}). Computing the $\frac{\eta}{s}$ ratio for this family provided two possible values to $\beta$. The first one, $\beta=1$, was physically expected, corresponding to the AdS$_5$--Schwarzschild   black brane. Besides the importance of the result itself, in particular for the membrane paradigm of AdS/CFT, it has a good potential for relevant applications, mainly in AdS/QCD. Taking into account the thermodynamics that underlies the family of deformed black branes solutions, arising from the Einstein--Hilbert action in the bulk, with a Gibbons--Hawking term and a counter-term that eliminates divergences, yields 
the deformed black brane temperature (\ref{Tofr}). This expression, \pptx{together with the fact that  the event horizon of the deformed AdS$_5$--Schwarzschild   black brane must assume real values,} constrain the range of the free parameter $\beta$ in the range (\ref{ra2}). 

\bltx{ Although we have derived our results using the ADM formalism, in a bulk embedding scenario, the KSS bound violation in the range $1<\beta\leq 2$ represents, as a matter of speculation, a possible smoking gun towards the fact that the deformed AdS$_5$--Schwarzschild black brane (\ref{1}), with metric coefficients (\ref{eq:Nu}, \ref{eq:Au}), 
might  be, alternatively, derived from an action with higher curvature terms. However, up to our knowledge, no result has been obtained in this aspect, yet.}

\bltx{The  family of AdS$_5$--Schwarzschild deformed black branes,  here derived using the ADM formalism, is also not the first example in the literature of a setup that violates the KSS bound \emph{and} does not involve higher derivative theories of gravity, in the   gauge/gravity correspondence. In fact, strongly coupled $\mathcal{N} = 4$ super-Yang--Mills plasmas can describe pre-equilibrium stages of the quark-gluon plasma (QGP) in heavy-ion collisions. In this setup, the shear viscosity, transverse to the direction of anisotropy, was shown to  saturate the KSS viscosity bound \cite{Rebhan:2011vd}. Besides, anisotropy in the shear viscosity induced by external magnetic fields in a strongly coupled plasma also provided violation in the KSS bound \cite{Critelli:2014kra}. Theories with higher order curvature terms in the action, in general, comprise attempts of describing quantum gravity.
Hence, one is restricted to consider CFT for which  the central charges  satisfy $|c - a|/c \ll 1$ and $c>a$, in such a way that still $c\sim a \gg 1$, also yielding violation of the KSS bound \cite{Buchel:2008vz,viol1}.   Up to now,  the equations of motion for 5D actions with higher curvature terms up to third order are already established in the literature, but it has been not possible to obtain  the deformed AdS$_5$--Schwarzschild black brane (\ref{1}) yet as an exact solution to any of them. We keep trying to compute higher curvature terms, including  fourth order terms, and we have not exhausted all the possibilities, yet.  Any effective action is expected to contain curvature terms of  higher  order, each one of them accompanying their respective coefficients. To derive a sensible derivative expansion, one should restrict to the classes of CFTs wherein these coefficients are proportional to inverse powers of the central charge $c$ \cite{Buchel:2008vz}.}

As large-$N_c$ gauge theories considered by AdS/CFT are good approximations to QCD, one could expect that the result of Eq. \eqref{eq:eta_s_sads_5} may be applied to the QGP, which is a natural phenomenon in QCD, when at high enough temperature the quarks and gluons are deconfined from protons and neutrons to form the QGP  \cite{qgpexp2}. In fact, experiments in the RHIC have shown that the QGP behaves like a viscous fluid with very small viscosity, which implies that the QGP is strongly-coupled, thus discarding the possibility of using perturbative QCD to the study of the plasma. Therefore, the new AdS$_5$--Schwarzschild deformed  black brane  (\ref{1}) can be widely used to probe additional properties in the AdS/QCD approach.  As in  the holographic soft-wall AdS/QCD the AdS$_5$-Schwarzschild black brane provides a reasonable description of mesons at finite temperature \cite{BoschiFilho:2004ci}, we can test if using the AdS$_5$--Schwarzschild deformed  black brane derives a more reliable meson mass spectra for the mesonic states and their resonances, better matching experimental results. Besides,  the new AdS$_5$--Schwarzschild deformed  black brane can be also explored in the context of the Hawking--Page transition and information entropy \cite{Bernardini:2016hvx,Braga:2016wzx}.

\section*{Acknowledgements}
\textcolor{black}{AJFM} is grateful to FAPESP (Grants No.  2017/13046-0  and No.  2018/00570-5) and to CAPES - Brazil. 
The work of PM was financed  in part by the Coordena\c c\~ao de Aperfei\c coamento de Pessoal de N\'ivel Superior -- Brasil (CAPES) -- Finance Code 001. 
 RdR~is grateful to FAPESP (Grant No.  2017/18897-8), to the National Council for Scientific and Technological Development  -- CNPq (Grants No. 303390/2019-0, No. 406134/2018-9 and No. 303293/2015-2), and to ICTP, for partial financial support.
\begin{widetext}
\appendix\section{}
\label{app}
{{\begin{eqnarray}\label{r10}
f(r, r_0, \beta)&=&-\frac{1}{r^{10}}\Bigg\{-\left(10 (\beta -1)+r^6-3 r^2r_0^4\right) \left(\beta +r^6-r^2
   r_0^4-1\right)+\frac{4 r^8 \left(-2 \beta +r^6+r^2
   r_0^4+2\right)^2}{\left(\beta +r^6-r^2 r_0^4-1\right)^2}\nonumber\\
  && +\frac{4 r^8 \left(4 r^{12}\!+\!8 (2\!-\!3 \beta ) r^8 r_0^4\!+\!(20 \beta \!-\!23) r^4
   r_0^8\!+\!3 (4 \beta \!-\!1) r_0^{12}\right)^2}{\left(2 r^8-5 r^4 r_0^4+3 r_0^8\right)^2 \left(2
   r^4+(1-4 \beta ) r_0^4\right)^2}\nonumber\\
  && -\frac{2 r^8 \left(8 r^{16}-60
   r^{12} r_0^4+6 (40 \beta  (2 \beta -3)+67) r^8 r_0^8+(4 \beta -1) (20 \beta +43) r^4
   r_0^{12}-9 (1-4 \beta )^2 r_0^{16}\right)}{\left(2 r^8-5 r^4 r_0^4+3 r_0^8\right) \left(2
   r^4+(1-4 \beta ) r_0^4\right)^2}\nonumber\\&&+\frac{1}{2
   r^4\!+\!(1\!-\!4 \beta ) r_0^4}[r^2 \left(2 r^8\!+\!2 r^6\!-\!5 r^4
   r_0^4\!+\!(1\!-\!4 \beta ) r^2 r_0^4\!+\!3 r_0^8\right) \left(\beta \!+\!r^6\!-\!r^4\!-\!r^2 r_0^4\!-\!1\right)]\nonumber\\&&+\!\frac{4 r^8 \left(r^6\!+\!r^2 r_0^4\!+\!2\!-\!2\beta\right) \left(4
   r^{12}\!+\!8 (2\!-\!3 \beta ) r^8 r_0^4\!+\!3 (4 \beta \!-\!1)
   r_0^{12}\right)}{\left(2 r^4\!-\!3 r_0^4\right) \left(r^4\!-\!r_0^4\right) \left(2 r^4\!+\!(1\!-\!4 \beta )
   r_0^4\right) \left(\beta\!+\!r^6\!-\!r^2 r_0^4\!-\!1\right)}\nonumber\\
  &&  
   +2 r^8 \left(\frac{2 r^8+5 r^4 r_0^4-9 r_0^8}{2 r^8-5 r^4 r_0^4+3 r_0^8}-\frac{4 r^4}{2 r^4+(1-4 \beta )
   r_0^4}+\frac{r^2 \left(3
   r^4-r_0^4\right)}{\beta +r^6-r^2 r_0^4-1}\right)\Bigg\}
   \end{eqnarray}}}
\section{} \label{appB}
\par We will show that the graviton propagates at the speed of light. Throughout this appendix we make $r_0=1$, and the metric \eqref{1} is written in coordinates $\{t,r,x,y,z\}$, where $r=u^{-1}$ according to the present convention.

\par Consider a perturbation of the form \eqref{ref12}. As discussed in the text, the perturbation $h_{xy}$ can be considered as a field on its own, hence we define $\varphi=g^{xx}h_{xy}$. We now identify the action as $S\sim S_0+S_2$, where $S_0$ does not have any contribution from $\varphi$, i.e. it is the action as studied in Sect. \ref{termo}, whereas $S_2$ contains contributions of $\varphi$ and its derivatives. Let
\begin{equation} \label{ap:pertbulk}
\varphi=\int dk\Phi(r)e^{-i\omega t+ikr+iqz}\ ,
\end{equation}
where $dk=\frac{d\omega dq dk}{\left(2\pi\right)^{3}}$, so that $S_2\propto \int\mathcal{L}(\Phi,\Phi^{\prime},\Phi^{\prime \prime})$, the proportionality factor is discarded. The Lagrangian reads
\begin{align}
\begin{aligned} \label{eqB2}
\sqrt{\frac{A}{N}}\mathcal{L}&=\Phi^{2}\left[-6r^{5}+2rA+r^{2}\left(ikA+2A^{\prime}+\frac{2AN^{\prime}}{N}\right)+\frac{7}{2}r^{3}\left(-q^{2}-Ak^{2}+\frac{\omega^{2}}{N}\right)+r^{3}\left(ikA^{\prime}+ikA\frac{N^{\prime}}{N}+\frac{AN^{\prime\prime}}{2N}\right.\right.\\&\left.\left.+\frac{A^{\prime}N^{\prime}}{4N}-\frac{AN^{\prime2}}{4N^{2}}\right)\right]+\Phi\Phi^{\prime}\left[r^{2}\left(8+7ikr\right)A+A^{\prime}+\frac{AN^{\prime}}{N}\right]+\frac{3}{2}\Phi^{\prime2}A+2A\Phi\Phi^{\prime\prime}\ .
\end{aligned}
\end{align}
For an action dependent on a single field up to its second derivative one can show immediately that
\begin{equation}\label{eqB3}
	\delta S=\delta S_{bdy}+\int dr\delta\Phi\left[\left(\frac{\partial\mathcal{L}}{\partial\Phi^{\prime\prime}}\right)^{\prime\prime}-\left(\frac{\partial\mathcal{L}}{\partial\Phi^{\prime}}\right)^{\prime}+\frac{\partial\mathcal{L}}{\partial\Phi}\right]\ ,
\end{equation}
$\delta S_{bdy}$ are surface terms, while the factor inside the integral is the equation of motion.

\par The momentum vector is $k^{\mu}=(\omega,k,0,0,q)$. Evaluating the EOM from \eqref{eqB3} using Lagrangian \eqref{eqB2} we obtain, in the limit $k^{\mu}\mapsto\infty$, the following
\begin{equation}
	k_{\mu}k^{\mu}=0\ ,
\end{equation}
i.e. the EOM for a light-ray. This shows that the graviton -- field associated to the perturbation \eqref{ref12} -- propagates with the speed of light.
\end{widetext}
\bibliography{bibliografia}
\bibliographystyle{iopart-num}

\end{document}